\documentclass[aps,8pt,twocolumn,tightenlines,floatfix]{revtex4}
\usepackage[dvips]{graphicx}
\usepackage[english]{babel}
\usepackage{amsmath}
\usepackage{amssymb}
\usepackage{slashed}

\begin{document}

\title{Transport Anomalies Associated with the Pseudogap From
a Preformed Pair Perspective}

\author{Vivek Mishra*}
\author{Dan Wulin*}
\author{K. Levin}
\affiliation{James Franck Institute and Department of Physics,
University of Chicago, Chicago, Illinois 60637, USA}
\altaffiliation{These two authors contributed equally to this work.}

\begin{abstract}
Transport
studies seem to be one of the strongest lines of support for
a preformed pair approach to the pseudogap. In
this paper we provide a fresh,
physically transparent look
at two important quantities: the diamagnetic susceptibility and conductivity.
We use a three dimensional preformed pair framework which has had
some success in the cold Fermi gases and
in the process
we reconcile recently
observed inconsistencies.
Specifically, while the preformed pairs in our theory give a
large contribution to the diamagnetic susceptibility, the
imaginary part of the conductivity is suppressed to zero much closer to $T_c$, as
is observed experimentally.
\end{abstract}

\maketitle

One of the biggest challenges in understanding the high temperature
superconductors revolves around the origin of the ubiquitous pseudogap.
Because this normal state gap has $d$-wave like features compatible with the superconducting
order parameter, this
suggests that the pseudogap
is related to some form of 
``precursor pairing" which would generalize
the behavior in conventional BCS superconductors, 
(where pairing and condensation 
take place at precisely the same temperature).
On the otherhand, there are many reports \cite{Taillefer4,Hinkov} suggesting
that the pseudogap onset temperature is
associated with a broken symmetry and, thus, another order parameter. 
It is widely believed that because the pseudogap has clear
signatures in generalized transport,
these measurements
may help with the centrally important question of
distinguishing the two scenarios. 
In this paper we provide a fresh, transparent look at
transport in the presence of a pseudogap where the
latter is associated with pre-formed pairs deriving from
a stronger than BCS attractive interaction. 
We are
thereby able to reconcile
inconsistencies
with cuprate experiments.
Importantly, there is no more theoretical flexibility
here than
in standard BCS theory so that predictions are concrete and
testable.

Our goal is to address the observed conflict between transport experiments 
\cite{Bilbro,Bilbro2}  
and a variety of precursor superconductivity scenarios before
reaching the definitive conclusion that the pseudogap 
derives from a non-superconducting order parameter. We argue here that
it is necessary to investigate one more precursor superconductivity
approach. Most importantly, this particular scenario, based on a stronger than
BCS attraction, has been
realized experimentally-- 
in atomic Fermi gases \cite{Ourviscosity} which
also appear to exhibit a pseudogap
\cite{Jin6,Feld2011,CSTL05}. We argue it
should also be applicable to those superconductors (such as the cuprates) with
anomalously high pairing onset temperature $T^*$, and
small
pair size.
Similar ideas were introduced by Geshkenbein, Ioffe
and Larkin \cite{Geshkenbein1997}.
In contrast to previous work here we discuss transport both above and
below the transition $T_c$ and we pay central attention to
the important
conductivity sum rule constraint. In view of the
strong evidence for three dimensional (d) 
critical behavior \cite{Kamal,Overend,Pureur}
we do not restrict consideration to strictly 2d systems.

The inconsistencies which we aim to reconcile pertain to the behavior
of the complex conductivity $\sigma = \sigma_1 + i \sigma_2$ and the
diamagnetic susceptiblity $\chi$.
[The widely discussed
Nernst effect was examined in earlier work \cite{TanNernst}.]
Below the transition, $\sigma_2$ directly relates to the superfluid
density. If, above $T_c$,
$\sigma_2$
were interpreted to reflect a remnant of the
superfluid density
(as expected in a simple fluctuation (\cite{LarkinVarlamov} or
more mesoscopic phase fluctuation theory \cite{Hanke,Behnia}),
this would suggest a close relationship between $\sigma_2$ and the
normal state diamagnetic susceptibility $\chi$, which is not observed
\cite{Bilbro}.
Problematic for a slightly different precursor scenario (the normal state vortex picture \cite{Ong2})
is the unexpectedly small
(by two orders of magnitude \cite{Bilbro2}) value of the ratio of the
real part of the conductivity $\sigma_1$
to $\chi$ in the normal state.

\begin{figure*} \includegraphics[width=6.5in,clip] {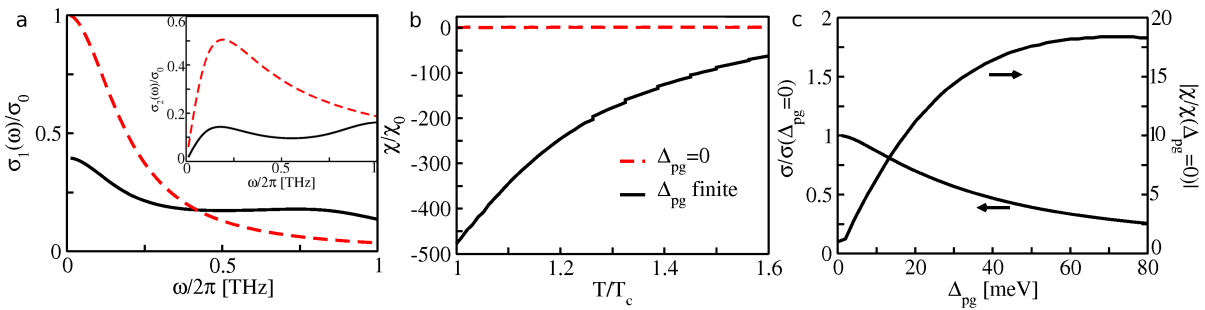}
\caption{
 Schematic figures showing the effect for $ T > T_c$
of the pseudogap on $\sigma(\omega)$ and
$\chi^{\textrm{dia}}$. (a) The real part of the conductivity $\sigma_1$ as a function of
frequency with (black curve) and without (red dashed curve) the pseudogap. Inset: The
imaginary part of the conductivity $\sigma_2$ as a function of frequency with (black
curve) and without (red dashed curve) the pseudogap.
(b) The diamagnetism with (black
curve) and without (red dashed curve) the pseudogap.
(c) The conductivity and diamagnetic
susceptibility as functions of $\Delta_{pg}$ at $T=15meV$. } \label{fig:1}
\end{figure*}

Our physical picture for the 
way in which transport is affected by preformed pairs
is relatively simple to understand.
In the presence of stronger 
than BCS attraction
there are both fermionic and metastable
Cooper pair degrees of freedom. The latter can be viewed as non-condensed
pairs, or pair-correlated fermions. It can be seen from simple
Boltzmann arguments \cite{LarkinVarlamov} that bosons
provide very large transport responses, provided they are
in 
proximity to
condensation. The 
Bose-Einstein distribution function
which is then peaked at small wavevectors, is in stark
contrast to its fermionic, Pauli principle
restricted counterpart; it leads to
a much stronger bosonic response
to
external field perturbations.
Importantly, if one associates the pseudogap
with long lived and meta-stable pairs in three dimensional
systems, these enhancements, in transport
can be shown to persist \cite{TanNernst} to temperatures
nearer to $T^* >>T_c$, as
one sees in a variety of
different transport experiments.
This should be distinguished from conventional fluctuation
effects \cite{Behnia}, which contribute in the critical
regime very close to $T_c$.

In the usual BCS-like, purely fermionic Hamiltonian only fermions
possess a hopping kinetic energy and
and thereby directly
contribute to transport. The contribution to transport from
pair correlated fermions 
enters 
indirectly by liberating these fermions through a break-up of the pairs.
Technically, we can associate this coupling to fermionic
transport as via the well known Aslamazov-Larkin diagram, importantly
modified to include the self consistently determined fermionic pairing gap.

A stronger than BCS attractive interaction
can be accomodated by a simple extension of Gor'kov theory.
This leads to non-condensed pair effects \cite{CSTL05}
above and below $T_c$.
Important here is the general form of the superconducting
electromagnetic response which
consists of three distinct contributions:  
(1) superfluid acceleration, 
(2) quasi-particle
scattering   
and (3) pair breaking and pair forming. 
These all appear in conventional BCS superconductors,
but at $T = 0$ this last effect is only present when there
is disorder.
However, in the presence of stronger than BCS attraction,
and at $T \neq 0$, non-condensed pairs provide an alternate way to decrease the
superfluid density, 
and the pair breaking and pair forming contributions will be
concomitantly more prominent \cite{Arcstransport}.

Without any detailed calculations we are now in a position to
predict results associated with the Thz conductivity
and the diamagnetism, which will be supported by later microscopic theory.
We now show how $\sigma_1(\omega \approx 0)$ is depressed by
the presence of a pseudogap $\Delta_{pg}$, how 
$\sigma_2(\omega)$ over a range of $\omega$ is also
depressed while $\chi$ is greatly enhanced.

Fig.\ref{fig:1}(a) shows how the normal state $\sigma_1(\omega)$
(and in the inset $\sigma_2(\omega)$) behaves
as a function of frequency. The red dashed curves are the results of
conventional Drude theory.
What happens when an above $T_c$ pseudogap is present is shown by
the black curves. The curves are normalized by $\sigma_0$, the 
normal state value of
the conductivity at $\omega=0$ in Drude theory.
Both theories (with or without the pseudogap) are consistent
with the f-sum rule, and thus have the same
fermionic carrier number
$\Big(\frac{n}{m} \Big)_{xx}$.
In the dc regime, with a pseudogap present, there are fewer fermions available to contribute to
transport. Their number is reduced by the pseudogap.
However, once the frequency is sufficient
to break the pairs into individual fermions,
the conductivity rises above that of the Drude model.
One can see that the effect of the pseudogap is to transfer the spectral
weight from low frequencies to higher energies ($\omega \approx 2 \Delta$,
where $\Delta$ is the pairing gap, and $\Delta \equiv \Delta_{pg}$ above $T_c$).
In this way one finds an extra ``mid-infrared" contribution to the
conductivity which is, as observed \cite{AndoRes1}
strongly tied to the presence of a pseudogap.

The behavior of
$\sigma_2(\omega)$, shown in the inset, is rather similarly
constrained.
On general principles, $\sigma_2$ must vanish at strictly zero frequency - as long
as the system is normal. Thus both
the red and black curves show that $\sigma_2(\omega \equiv 0) = 0$.
Here one can see that the low frequency behavior is also suppressed by the
presence of a pseudogap because of the gap-induced decrease in the number of carriers.
Similarly, the second peak (around $2 \Delta$) in $\sigma_1(\omega)$
leads, via a Kramers Kronig transform
to a slight depression in $\sigma_2(\omega) $ in this frequency range.
Hence as shown in the inset, $\sigma_2(\omega)$ is significally reduced
relative to the Drude result and tends overall
to increase with $\omega$. There is virtually no sign of
a $\omega^{-1}$ upturn in $\sigma_2$ which would reflect a
remnant of the superfluid density above $T_c$. This presumably
is a fluctuation effect which pertains to the narrow critical
regime.

In Fig.\ref{fig:1}(b) we present similar comparisons of the behavior
of the orbital susceptibility above $T_c$ in a non-gapped
normal state (red dashed curve) and
in the presence of a pseudogap (black curve). The curves are normalized by $\chi_0$, the absolute value of the diamagnetic susceptibility for $\Delta_{pg}=0$ at $T=T_c$.
One can see that in the absence of a pseudogap only a
very weak Landau diamagnetism appears.
However, the figure shows that in the presence of
a pseudogap the diamagnetic contribution is significantly enhanced.
This diamagnetism originates from the large electromagnetic response
associated with bosonic degrees of freedom;
the breaking of pairs allows this diamagnetism to be reflected in
the fermionic response.
It should be noted that Van Hove effects
enhance this diamagnetism,
as does $d$-wave pairing which leads to
an excess of low energy fermionic excitations. Moreover, this diamagnetism is
not restricted to two dimensional models.
       
All of this leads to a simple anti-correlation between the dc conductivity and the diamagnetic susceptibilty in the normal state,
which is shown in
Fig.\ref{fig:1}(c).
Here we plot on the left and right hand axes the zero frequency conductivity as
a function of varying pseudogap energy scale $\Delta_{pg}$ and the orbital
(diamagnetic) susceptibility with varying $\Delta_{pg}$ respectively.
The former is depressed as the pairing gap increases whereas the latter is enhanced.

\begin{figure*}
\includegraphics[width=6.25in,clip]
{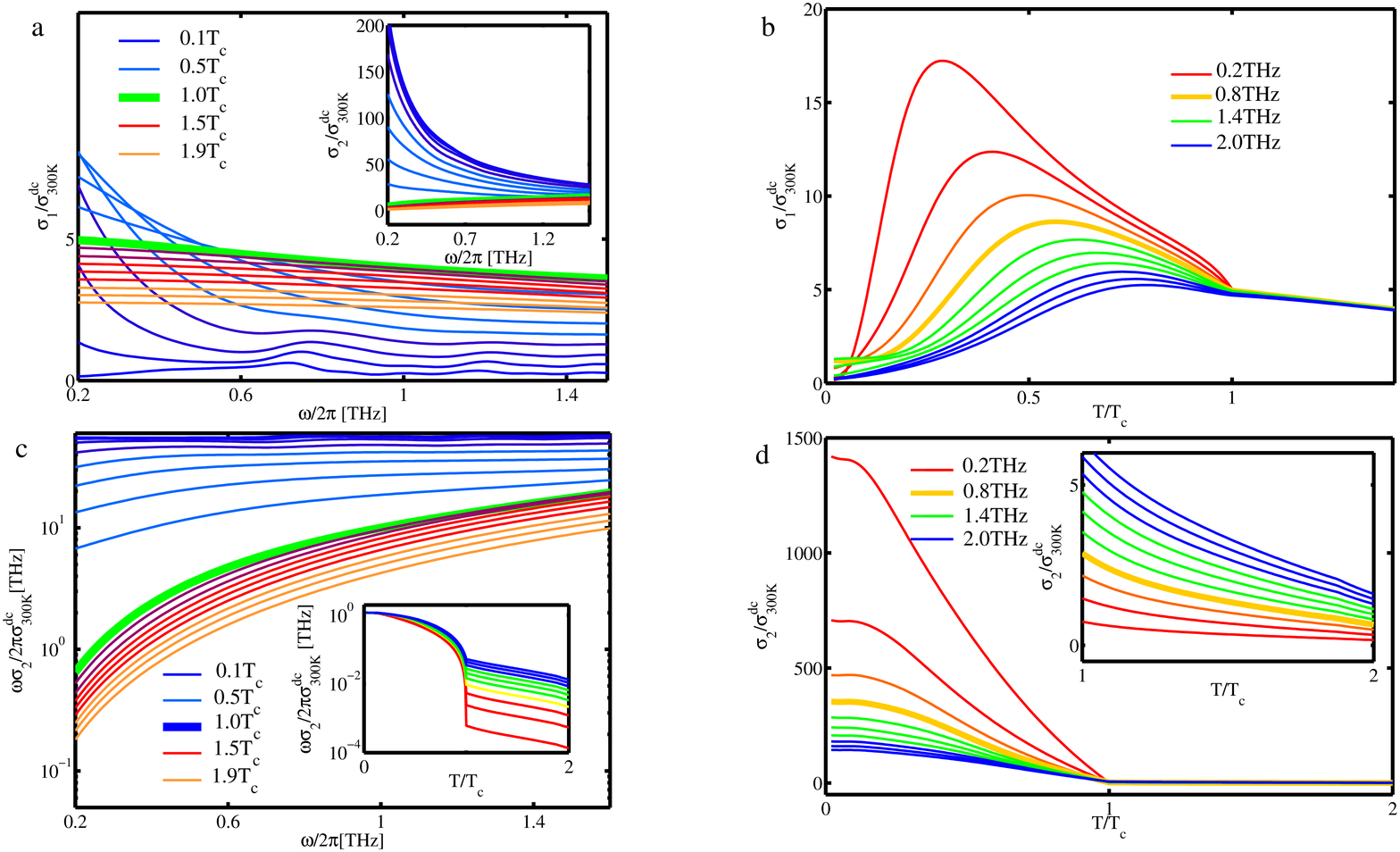}
\caption{Results for $\sigma_1$ and $\sigma_2$. (a) $\sigma_1$ as a function of frequency. Inset: $\sigma_2$ as a function of frequency. 
(b)$\sigma_1$ as a function of temperature. 
(c) $\omega\sigma_2$ as a function of frequency. Inset: $\omega\sigma_2$ as a function of temperature.
(d) $\sigma_2$ as a function of temperature. Inset: $\sigma_2$
as a function of temperature near $T_c$.} 
\label{fig:2}
\end{figure*}

These same conclusions (which are qualitatively compatible
with experiment \cite{Ong2,Bilbro,Bilbro2} derive from microscopic theory. 
Here the linear response of the 
electromagnetic current $\textbf{J}$ to a small 
vector potential $\textbf{A}$ is characterized by the  
tensor $\tensor{P}+\tensor{n}/m$ through the equation 
$\textbf{J}=-(\tensor{P}+\tensor{n}/m)\textbf{A}$. 
The transverse f-sum rule is
an important constraint on any theory of transport
\begin{eqnarray} 
\displaystyle{\lim_{\textbf{q}\rightarrow0}}\displaystyle{\int_{-\infty}^{\infty}}\frac{d\omega}{\pi}\Big(-\frac{\textrm{Im}P_{xx}(\textbf{q},\omega)}{\omega}\Big)=\Big(\frac{n_n}{m} \Big)_{xx}
\label{eq:sumrule} 
\end{eqnarray} where $\tensor{n_n}/m$ is the normal fluid density and $P_{xx}$ is the diagonal 
component of the paramagnetic current, $\tensor{P}$, along the x-direction. 
Similarly, $(n/m)_{xx}$ is the diagonal component of $\tensor{n}/m$ along the x-direction.
We stress that
only the fermionic density (and mass) appears on the
right hand side of Eq.~(\ref{eq:sumrule}).
The sum rule establishes a strong connection between transport and
the fermionic kinetic energy, so that many body interactions only serve to
redistribute the spectral weight.
Thus, for example, even
though meta-stable pairs are present, their contribution to transport is
indirect and appears when such pairs can be decomposed.
This version of the f-sum
rule applies to any many body Hamiltonian which contains an arbitrary two body
interaction and a kinetic energy associated with fermions.

Throughout, we work in the transverse gauge.
As a consequence all effects of the order parameter collective modes
(which are longitudinal) do not enter.
The complex 
conductivity is microscopically defined in terms of $\tensor{P}$ and $\tensor{n}/m$:
 \begin{eqnarray} 
\sigma(\omega)=-\displaystyle{\lim_{\textbf{q}\rightarrow0}}\frac{P_{xx}(\textbf{q},\omega)+(n/m)_{xx}}{i\omega}\label{eq:conddef} 
\end{eqnarray}
Above $T_c$, the linear diamagnetic response is similarly related to $P_{xx}$ and $(n/m)_{xx}$. It is given 
by

\begin{equation}
\chi^{\textrm{dia}}_{zz}=-\displaystyle{\lim_{q_{y}\rightarrow0}}\Bigg[\frac{
\textrm{Re}P_{xx}(\textbf{q},\omega=0)+(n/m)_{xx}}{{q}_{y}^2}\label{eq:diadef}
\Bigg]_{q_x=q_z=0}
\end{equation}
In the superfluid phase the tensors $\tensor{P}$ and $\tensor{n}/m$ no longer
cancel when 
$\textbf{q}\rightarrow0$, reflecting the Meissner effect.
We stress that Equations \eqref{eq:sumrule}-\eqref{eq:diadef} are completely general.

\begin{figure}
\includegraphics[width=3.in,clip]
{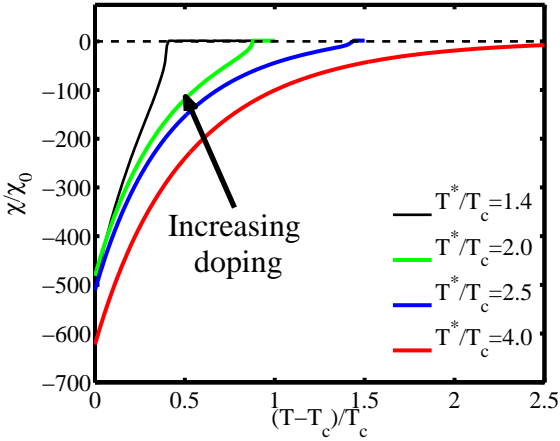}
\caption{
 The diamagnetism as a function of temperature for different ``hole
concentrations" as parameterized via the indicated $T^*/T_c$. }
\label{fig:3}
\end{figure}

We now turn to more microscopic calculations.
Previous papers \cite{CSTL05,Chen2} have described how the parameters $\Delta(T)$,
$\Delta_{pg}(T)$, and $\mu$ are self consistently obtained and
how one accomodates a variety of dopings, by effectively fitting the attractive
interaction to match $T^*$ and $T_c$. Our figures correspond
to moderate underdoping. A nearest
neighbor tightbinding dispersion
$\xi_{\textbf{p}}=-2t[\textrm{cos}(k_xa)+\textrm{cos}(k_ya)]-\mu$ with $t=300 meV$ is
used throughout, and for simplicity, we took a simple $\tau \propto T^{-2}$ power law
(associated with the Fermi arcs \cite{Kanigelarcs})
for the transport lifetime. Very few of our results depended on this assumption
which was made in earlier work \cite{Arcstransport}.
A general finding is that, while $\Delta_{pg}$ decreases monotonically from $T_c$ to
$T^*$, with decreasing $T$ below $T_c$, $\Delta_{pg}$ decreases while
$\Delta_{sc}$ rises, reflecting the fact
that finite momentum pairs are converted to the $q = 0$ condensate,
while maintaining an overall nearly constant $\Delta(T)$.
We have previously derived
microscopic representations of $\tensor{P}$ and $\tensor{n}/m$ \cite{Kosztin2,Chen2,Arcstransport}.
Importantly, our
gauge invariant electromagnetic response function analytically satisfies the transverse f-sum rule.
One can derive these contributions in a variety of ways but the most straightforward
involves inclusion of generalized Maki-Thompson and Aslamazov-Larkin diagrams. The latter
are considered to be effectively equivalent to
Boltzmann (or time dependent Ginsburg-Landau-like) approaches to bosonic transport.
Here the stronger-than-BCS attraction enters in an important way in order to insure
that the pairing gap energy scale $\Delta$ is explicitly incorporated.
The paramagnetic
tensor current-current correlation function $\tensor{P}$
is

\begin{widetext}
 \begin{eqnarray}
{P}_{xx}(\textbf{q},\omega)=\displaystyle{\sum_{\textbf{p}}}\frac{\partial\xi_{
\text{\bf
p}}}{\partial p_{x}}\frac{\partial\xi_{\text{\bf p}}}{\partial
p_{x}}\Bigg[\frac{E^++E^-}{E^+E^-}\frac{E^+E^--\xi^+\xi^--\Delta_{sc}^2+\Delta_
{pg}^2}{\omega^2-(E^++E^-)^2}\Big(1-f(E^+)-f(E^-)\Big)\nonumber\\-\frac{E^+-E^-
}{E^+E^-}\frac{E^+E^-+\xi^+\xi^-+\Delta_{sc}^2-\Delta_{pg}^2}{\omega^2-(E^+-E^-
)^2}\Big(f(E^+)-f(E^-)\Big)\Bigg]\label{eq:ptensor}
\end{eqnarray}
\end{widetext}
where $\omega$ has a small imaginary part and $f$ the Fermi function. Here
$E_{\mathbf{p}} \equiv \sqrt{ \xi_{\mathbf{p}}^2 + \Delta^2(T)  }$, where $\xi_{\mathbf{p}}
=\epsilon_{\mathbf{p}}-\mu$,
and
$\Delta_{sc}$
($\Delta_{pg}$) is the gap component of the condensed (non-condensed)
pairs,
with $\Delta=\sqrt{\Delta_{sc}^2+\Delta_{pg}^2}$.
All transport
expressions in this paper reduce to those of strict BCS theory
when the attraction is weak and $\Delta_{\textrm{pg}} =0$.
Here we
define $E^{\pm}=E_{\textbf{p}\pm\textbf {q}/2}$ and $\xi^{\pm}=\xi_{\textbf{p}\pm\textbf{q}/2}$.
Importantly, the terms on the first line in
Eq.\eqref{eq:ptensor} represent the pair breaking and pair forming contributions.
The second line is associated with fermionic scattering.
Also important to the electromagnetic response is the number density
$\tensor{n}/m$ which can be rewritten as
\begin{eqnarray}
=2\sum_{\text{\bf p}}\frac{\partial\xi_{\text{\bf p}}}{\partial
p_{x}}\frac{\partial\xi_{\text{\bf p}}}{\partial p_{x}}\Bigg[
\frac{\Delta^{2}_{sc}\!+\!\Delta^{2}_{pg}}{E_{p}^{2}}\left(\frac{1\!-\!2
f(E_{p})}{2 E_{p}}\right)
\!-\! f^{\prime}(E_{p})  \Bigg].
\label{eq:ntensor}
\end{eqnarray}

Eqs.\eqref{eq:conddef}-\eqref{eq:diadef} 
yield analytic expressions for $\sigma(\omega)$ 
and $\chi^{\textrm{dia}}$. 
Fig.\ref{fig:2} displays our more
quantitative results for $\sigma_1$ and $\sigma_2$ as both functions of 
$\omega$ and $T$. The layout is designed to duplicate figures from Ref. \onlinecite{Bilbro}
and the general trends are similar. Thus one sees from Fig.\ref{fig:2}(a) and its inset
that well above $T_c$, the real part of the conductivity is almost
frequency independent. The imaginary part is small in this regime. At the lowest
temperatures $\sigma_1$ contains much reduced spectral weight while the
frequency dependence of $\sigma_2 \propto \omega^{-1}$; both of these reflect the
characteristic behavior 
of a superfluid.

Here as in the experimental studies \cite{Bilbro},
we focus primarily on the temperature dependent plots 
in Figs.\ref{fig:2}(b), (d) and the inset to (c).
One sees that $\sigma_1$ shows a slow decrease as the temperature is 
raised above $T_c$.
Somewhat below $T_c$, $\sigma_1$ exhibits a peak which occurs at progressively lower
temperatures as the probe frequency is decreased. Roughly at $T_c$ we find
that $\sigma_2$ shows a sharp upturn at low $\omega$. The region of finite $\sigma_2$
above the transition can be seen from the inset in 
Fig.\ref{fig:2}(d) and it 
is clearly very small in the pseudogap state. 
The inset of Fig.\ref{fig:2}(d) shows an expanded view of
$\sigma_2(T)$ near $T_c$. 
In agreement with
experiment, the nesting of the $\sigma_2$ versus T curves switches orders above
$T_c$.
This important point
reflects the fact that $\sigma_2(\omega)$ is generally increasing
with increasing $\omega$ above $T_c$
as seen in the inset in Fig.~\ref{fig:1}(a) and in experiment.
This is in contrast to the behavior expected of a fluctuation
contribution where a $\omega^{-1}$ dependence would occur.
However, in slightly different plots, the counterpart
experimental studies reveal a
small 10-15K range where this fluctuation contribution
is visible.
This effect would not be present in a mean field approach.
As speculated in Ref.~\onlinecite{Bilbro}, one should
distinguish these near $T_c$ critical fluctuations from
preformed pairs which persist to much higher temperatures.

These effects are made clearer by plotting the
``phase stiffness" which is proportional to the quantity
$\omega\sigma_2$   
and is shown in Fig.\ref{fig:2}(c). Deep in the superconducting state
there is no $\omega$ dependence to $\omega \sigma_2(\omega)$, while
at higher $T$ this dependence becomes apparent. 
In the inset to (c), the temperature
dependence of $\omega\sigma_2(T)$ is displayed.
We see that 
above $T_c$, 
$\omega \sigma_2$ is never strictly constant, as would
be expected from fluctuation contributions.
In experiment the onset of 
finite frequency spreading of the curves at $ T \leq T_c$ has been
attributed to Kosterlitz-Thouless physics \cite{Corson1999}.

Finally, we turn to the diamagnetic response. 
Figure~\ref{fig:3} shows $\chi^{\textrm{dia}}$ as a function of temperature for four 
different dopings. Independently of the particular parameters that are used, it is seen 
that the magnitude of
$\chi^{\textrm{dia}}$ is enhanced even at temperatures well above $T_c$.
We should not associate this diamagnetism with short range Meissner currents,
as might be appropriate to alternative phase fluctuation \cite{Hanke} 
or normal state vortex scenarios \cite{Bilbro2}. Rather here,
the diamagnetism arises from the large contribution of non-condensed pairs
which are in proximity to condensation \cite{LarkinVarlamov,TanNernst}. 
This has a similarity
to low d fluctuation effects, but arises in the 3d systems here from stronger than BCS
attraction, which stabilizes these pair degrees of freedom.
Since the kinetic energy ultimately resides in the fermionic system, it is
not surprising that we find
it is the pair breaking terms 
which
provide
the conduit for communicating enhanced bosonic transport
contributions to the
fermionic transport channel.
Because we are working at effectively zero magnetic
field, we have not addressed diamagnetism associated with
non-linear response, although this appears to be very
anomalous experimentally \cite{Ong2}. At the leading order
level we have shown here that there is a profound connection between the complex
conductivity and this orbital magnetism.
Importantly, while the preformed pairs in our theory give a 
large contribution to the diamagnetic susceptibility, as is observed experimentally, the 
imaginary part of the conductivity is suppressed to zero much closer to $T_c$, as
observed.

We end with the following observations.
Our theoretical approach,
has virtually no flexibility; it was set up \cite{Kosztin2,Chen2} before
there was much experimental interest in these transport measurements.
In accord with experiment, we find:
(i) that pseudogap effects
lead to an enhanced diamagnetism above $T_c$,
(ii) that the imaginary conductivity $\sigma_2(\omega)$
is reduced to zero in a very narrow range of $T$ above $T_c$, and (iii)
that the real conductivity
$\sigma_1(\omega \approx 0)$
is suppressed as the pseudogap becomes larger.
This last point is in turn associated with a transfer of
low $\omega$ conductivity spectral weight into the mid-infrared region.

\vskip 5mm
This work is supported by NSF-MRSEC Grant
0820054. We thank P. Scherpelz and A. A. Varlamov for useful
conversations.

\bibliographystyle{apsrev}

\begin{thebibliography}{23}
\expandafter\ifx\csname natexlab\endcsname\relax\def\natexlab#1{#1}\fi
\expandafter\ifx\csname bibnamefont\endcsname\relax
  \def\bibnamefont#1{#1}\fi
\expandafter\ifx\csname bibfnamefont\endcsname\relax
  \def\bibfnamefont#1{#1}\fi
\expandafter\ifx\csname citenamefont\endcsname\relax
  \def\citenamefont#1{#1}\fi
\expandafter\ifx\csname url\endcsname\relax
  \def\url#1{\texttt{#1}}\fi
\expandafter\ifx\csname urlprefix\endcsname\relax\def\urlprefix{URL }\fi
\providecommand{\bibinfo}[2]{#2}
\providecommand{\eprint}[2][]{\url{#2}}

\bibitem[{\citenamefont{Daou et~al.}(2010)\citenamefont{Daou, Chang, LeBoeuf,
  Cyr-Choiniere, Laliberte, Doiron-Leyraud, Ramshaw, Liang, Bonn, Hardy
  et~al.}}]{Taillefer4}
\bibinfo{author}{\bibfnamefont{R.}~\bibnamefont{Daou}},
  \bibinfo{author}{\bibfnamefont{J.}~\bibnamefont{Chang}},
  \bibinfo{author}{\bibfnamefont{D.}~\bibnamefont{LeBoeuf}},
  \bibinfo{author}{\bibfnamefont{O.}~\bibnamefont{Cyr-Choiniere}},
  \bibinfo{author}{\bibfnamefont{F.}~\bibnamefont{Laliberte}},
  \bibinfo{author}{\bibfnamefont{N.}~\bibnamefont{Doiron-Leyraud}},
  \bibinfo{author}{\bibfnamefont{B.~J.} \bibnamefont{Ramshaw}},
  \bibinfo{author}{\bibfnamefont{R.}~\bibnamefont{Liang}},
  \bibinfo{author}{\bibfnamefont{D.~A.} \bibnamefont{Bonn}},
  \bibinfo{author}{\bibfnamefont{W.}~\bibnamefont{Hardy}},
  \bibnamefont{et~al.}, \bibinfo{journal}{Nature}
  \textbf{\bibinfo{volume}{463}}, \bibinfo{pages}{519} (\bibinfo{year}{2010}).

\bibitem[{\citenamefont{{Hinkov} et~al.}(2007)\citenamefont{{Hinkov},
  {Bourges}, {Pailhes}, {Sidis}, {Ivanov}, {Frost}, {Perring}, {Lin}, {Chen},
  and {Keimer}}}]{Hinkov}
\bibinfo{author}{\bibfnamefont{V.}~\bibnamefont{{Hinkov}}},
  \bibinfo{author}{\bibfnamefont{P.}~\bibnamefont{{Bourges}}},
  \bibinfo{author}{\bibfnamefont{S.}~\bibnamefont{{Pailhes}}},
  \bibinfo{author}{\bibfnamefont{Y.}~\bibnamefont{{Sidis}}},
  \bibinfo{author}{\bibfnamefont{A.}~\bibnamefont{{Ivanov}}},
  \bibinfo{author}{\bibfnamefont{C.~D.} \bibnamefont{{Frost}}},
  \bibinfo{author}{\bibfnamefont{T.~G.} \bibnamefont{{Perring}}},
  \bibinfo{author}{\bibfnamefont{C.~T.} \bibnamefont{{Lin}}},
  \bibinfo{author}{\bibfnamefont{D.~P.} \bibnamefont{{Chen}}},
  \bibnamefont{and} \bibinfo{author}{\bibfnamefont{B.}~\bibnamefont{{Keimer}}},
  \bibinfo{journal}{Nature Physics} \textbf{\bibinfo{volume}{3}},
  \bibinfo{pages}{780} (\bibinfo{year}{2007}).

\bibitem[{\citenamefont{Bilbro et~al.}(2011{\natexlab{a}})\citenamefont{Bilbro,
  Guilar, Logvenov, Pelleg, Bozovic, and Armitage}}]{Bilbro}
\bibinfo{author}{\bibfnamefont{L.~S.} \bibnamefont{Bilbro}},
  \bibinfo{author}{\bibfnamefont{R.~V.} \bibnamefont{Guilar}},
  \bibinfo{author}{\bibfnamefont{B.}~\bibnamefont{Logvenov}},
  \bibinfo{author}{\bibfnamefont{O.}~\bibnamefont{Pelleg}},
  \bibinfo{author}{\bibfnamefont{I.}~\bibnamefont{Bozovic}}, \bibnamefont{and}
  \bibinfo{author}{\bibfnamefont{N.~P.} \bibnamefont{Armitage}},
  \bibinfo{journal}{Nature Physics} \textbf{\bibinfo{volume}{7}},
  \bibinfo{pages}{2980302} (\bibinfo{year}{2011}{\natexlab{a}}).

\bibitem[{\citenamefont{Bilbro et~al.}(2011{\natexlab{b}})\citenamefont{Bilbro,
  Guilar, Logvenov, Bozovic, and Armitage}}]{Bilbro2}
\bibinfo{author}{\bibfnamefont{L.~S.} \bibnamefont{Bilbro}},
  \bibinfo{author}{\bibfnamefont{R.~V.} \bibnamefont{Guilar}},
  \bibinfo{author}{\bibfnamefont{B.}~\bibnamefont{Logvenov}},
  \bibinfo{author}{\bibfnamefont{I.}~\bibnamefont{Bozovic}}, \bibnamefont{and}
  \bibinfo{author}{\bibfnamefont{N.~P.} \bibnamefont{Armitage}},
  \bibinfo{journal}{Phys. Rev. B} \textbf{\bibinfo{volume}{84}},
  \bibinfo{pages}{100511(R)} (\bibinfo{year}{2011}{\natexlab{b}}).

\bibitem[{\citenamefont{Guo et~al.}(2011)\citenamefont{Guo, Wulin, Chien, and
  Levin}}]{Ourviscosity}
\bibinfo{author}{\bibfnamefont{H.}~\bibnamefont{Guo}},
  \bibinfo{author}{\bibfnamefont{D.}~\bibnamefont{Wulin}},
  \bibinfo{author}{\bibfnamefont{C.-C.} \bibnamefont{Chien}}, \bibnamefont{and}
  \bibinfo{author}{\bibfnamefont{K.}~\bibnamefont{Levin}},
  \bibinfo{journal}{Phys. Rev. Lett.} \textbf{\bibinfo{volume}{107}},
  \bibinfo{pages}{020403} (\bibinfo{year}{2011}).

\bibitem[{\citenamefont{Stewart et~al.}(2008)\citenamefont{Stewart, Gaebler,
  and Jin}}]{Jin6}
\bibinfo{author}{\bibfnamefont{J.~T.} \bibnamefont{Stewart}},
  \bibinfo{author}{\bibfnamefont{J.~P.} \bibnamefont{Gaebler}},
  \bibnamefont{and} \bibinfo{author}{\bibfnamefont{D.~S.} \bibnamefont{Jin}},
  \bibinfo{journal}{Nature} \textbf{\bibinfo{volume}{454}},
  \bibinfo{pages}{744} (\bibinfo{year}{2008}).

\bibitem[{\citenamefont{{Feld} et~al.}(2011)\citenamefont{{Feld},
  {Fr{\"o}hlich}, {Vogt}, {Koschorreck}, and {K{\"o}hl}}}]{Feld2011}
\bibinfo{author}{\bibfnamefont{M.}~\bibnamefont{{Feld}}},
  \bibinfo{author}{\bibfnamefont{B.}~\bibnamefont{{Fr{\"o}hlich}}},
  \bibinfo{author}{\bibfnamefont{E.}~\bibnamefont{{Vogt}}},
  \bibinfo{author}{\bibfnamefont{M.}~\bibnamefont{{Koschorreck}}},
  \bibnamefont{and}
  \bibinfo{author}{\bibfnamefont{M.}~\bibnamefont{{K{\"o}hl}}},
  \bibinfo{journal}{Nature} \textbf{\bibinfo{volume}{480}}, \bibinfo{pages}{75}
  (\bibinfo{year}{2011}).

\bibitem[{\citenamefont{Chen et~al.}(2005)\citenamefont{Chen, Stajic, Tan, and
  Levin}}]{CSTL05}
\bibinfo{author}{\bibfnamefont{Q.~J.} \bibnamefont{Chen}},
  \bibinfo{author}{\bibfnamefont{J.}~\bibnamefont{Stajic}},
  \bibinfo{author}{\bibfnamefont{S.}~\bibnamefont{Tan}}, \bibnamefont{and}
  \bibinfo{author}{\bibfnamefont{K.}~\bibnamefont{Levin}},
  \bibinfo{journal}{Phys. Rep.} \textbf{\bibinfo{volume}{412}},
  \bibinfo{pages}{1} (\bibinfo{year}{2005}).

\bibitem[{\citenamefont{Geshkenbein et~al.}(1997)\citenamefont{Geshkenbein,
  Ioffe, and Larkin}}]{Geshkenbein1997}
\bibinfo{author}{\bibfnamefont{V.~B.} \bibnamefont{Geshkenbein}},
  \bibinfo{author}{\bibfnamefont{L.~B.} \bibnamefont{Ioffe}}, \bibnamefont{and}
  \bibinfo{author}{\bibfnamefont{A.~I.} \bibnamefont{Larkin}},
  \bibinfo{journal}{Phys. Rev. B} \textbf{\bibinfo{volume}{55}},
  \bibinfo{pages}{3173} (\bibinfo{year}{1997}).

\bibitem[{\citenamefont{Kamal et~al.}(1994)\citenamefont{Kamal, Bonn,
  Goldenfeld, Hirschfeld, lian, and Hardy}}]{Kamal}
\bibinfo{author}{\bibfnamefont{S.}~\bibnamefont{Kamal}},
  \bibinfo{author}{\bibfnamefont{D.~A.} \bibnamefont{Bonn}},
  \bibinfo{author}{\bibfnamefont{N.}~\bibnamefont{Goldenfeld}},
  \bibinfo{author}{\bibfnamefont{P.~J.} \bibnamefont{Hirschfeld}},
  \bibinfo{author}{\bibfnamefont{R.}~\bibnamefont{lian}}, \bibnamefont{and}
  \bibinfo{author}{\bibfnamefont{W.~N.} \bibnamefont{Hardy}},
  \bibinfo{journal}{Phys. Rev. Lett.} \textbf{\bibinfo{volume}{73}},
  \bibinfo{pages}{1845} (\bibinfo{year}{1994}).

\bibitem[{\citenamefont{Overend et~al.}(1994)\citenamefont{Overend, Howson, and
  Lawrie}}]{Overend}
\bibinfo{author}{\bibfnamefont{N.}~\bibnamefont{Overend}},
  \bibinfo{author}{\bibfnamefont{M.}~\bibnamefont{Howson}}, \bibnamefont{and}
  \bibinfo{author}{\bibfnamefont{I.}~\bibnamefont{Lawrie}},
  \bibinfo{journal}{Phys. Rev. Lett.} \textbf{\bibinfo{volume}{72}},
  \bibinfo{pages}{3238} (\bibinfo{year}{1994}).

\bibitem[{\citenamefont{{Pureur} et~al.}(1993)\citenamefont{{Pureur},
  {Menegotto Costa}, {Rodrigues}, {Schaf}, and {Kunzler}}}]{Pureur}
\bibinfo{author}{\bibfnamefont{P.}~\bibnamefont{{Pureur}}},
  \bibinfo{author}{\bibfnamefont{R.}~\bibnamefont{{Menegotto Costa}}},
  \bibinfo{author}{\bibfnamefont{P.}~\bibnamefont{{Rodrigues}}},
  \bibinfo{author}{\bibfnamefont{J.}~\bibnamefont{{Schaf}}}, \bibnamefont{and}
  \bibinfo{author}{\bibfnamefont{J.~V.} \bibnamefont{{Kunzler}}},
  \bibinfo{journal}{Phys. Rev. B} \textbf{\bibinfo{volume}{47}},
  \bibinfo{pages}{11420} (\bibinfo{year}{1993}).

\bibitem[{\citenamefont{Tan and Levin}(2004)}]{TanNernst}
\bibinfo{author}{\bibfnamefont{S.}~\bibnamefont{Tan}} \bibnamefont{and}
  \bibinfo{author}{\bibfnamefont{K.}~\bibnamefont{Levin}},
  \bibinfo{journal}{Phys. Rev. B} \textbf{\bibinfo{volume}{69}},
  \bibinfo{pages}{064510} (\bibinfo{year}{2004}).

\bibitem[{\citenamefont{Larkin and Varlamov}(2005)}]{LarkinVarlamov}
\bibinfo{author}{\bibfnamefont{A.~I.} \bibnamefont{Larkin}} \bibnamefont{and}
  \bibinfo{author}{\bibfnamefont{A.~A.} \bibnamefont{Varlamov}},
  \emph{\bibinfo{title}{Theory of Fluctuations in Superconductors}}
  (\bibinfo{publisher}{Oxford University Press, New York},
  \bibinfo{year}{2005}).

\bibitem[{\citenamefont{Eckl and Hanke}(2006)}]{Hanke}
\bibinfo{author}{\bibfnamefont{T.}~\bibnamefont{Eckl}} \bibnamefont{and}
  \bibinfo{author}{\bibfnamefont{W.}~\bibnamefont{Hanke}},
  \bibinfo{journal}{Phys. Rev. B} \textbf{\bibinfo{volume}{74}},
  \bibinfo{pages}{134510} (\bibinfo{year}{2006}).

\bibitem[{\citenamefont{{Pourret} et~al.}(2006)\citenamefont{{Pourret},
  {Aubin}, {Lesueur}, {Marrache-Kikuchi}, {Berg{\'e}}, {Dumoulin}, and
  {Behnia}}}]{Behnia}
\bibinfo{author}{\bibfnamefont{A.}~\bibnamefont{{Pourret}}},
  \bibinfo{author}{\bibfnamefont{H.}~\bibnamefont{{Aubin}}},
  \bibinfo{author}{\bibfnamefont{J.}~\bibnamefont{{Lesueur}}},
  \bibinfo{author}{\bibfnamefont{C.~A.} \bibnamefont{{Marrache-Kikuchi}}},
  \bibinfo{author}{\bibfnamefont{L.}~\bibnamefont{{Berg{\'e}}}},
  \bibinfo{author}{\bibfnamefont{L.}~\bibnamefont{{Dumoulin}}},
  \bibnamefont{and} \bibinfo{author}{\bibfnamefont{K.}~\bibnamefont{{Behnia}}},
  \bibinfo{journal}{Nature Physics} \textbf{\bibinfo{volume}{2}},
  \bibinfo{pages}{683} (\bibinfo{year}{2006}).

\bibitem[{\citenamefont{Li et~al.}(2010)\citenamefont{Li, Wang, Komiya, Ono,
  Ando, Gu, and Ong}}]{Ong2}
\bibinfo{author}{\bibfnamefont{L.}~\bibnamefont{Li}},
  \bibinfo{author}{\bibfnamefont{Y.}~\bibnamefont{Wang}},
  \bibinfo{author}{\bibfnamefont{S.}~\bibnamefont{Komiya}},
  \bibinfo{author}{\bibfnamefont{S.}~\bibnamefont{Ono}},
  \bibinfo{author}{\bibfnamefont{Y.}~\bibnamefont{Ando}},
  \bibinfo{author}{\bibfnamefont{G.~G.} \bibnamefont{Gu}}, \bibnamefont{and}
  \bibinfo{author}{\bibfnamefont{N.~P.} \bibnamefont{Ong}},
  \bibinfo{journal}{Phys. Rev. B} \textbf{\bibinfo{volume}{81}},
  \bibinfo{pages}{054510} (\bibinfo{year}{2010}).

\bibitem[{\citenamefont{Wulin et~al.}(2011)\citenamefont{Wulin, Fregoso, Guo,
  Chien, and Levin}}]{Arcstransport}
\bibinfo{author}{\bibfnamefont{D.}~\bibnamefont{Wulin}},
  \bibinfo{author}{\bibfnamefont{B.}~\bibnamefont{Fregoso}},
  \bibinfo{author}{\bibfnamefont{H.}~\bibnamefont{Guo}},
  \bibinfo{author}{\bibfnamefont{C.-C.} \bibnamefont{Chien}}, \bibnamefont{and}
  \bibinfo{author}{\bibfnamefont{K.}~\bibnamefont{Levin}},
  \bibinfo{journal}{Phys. Rev. B} \textbf{\bibinfo{volume}{84}},
  \bibinfo{pages}{140509(R)} (\bibinfo{year}{2011}).

\bibitem[{\citenamefont{Lee et~al.}(2005)\citenamefont{Lee, Segawa, Li,
  Padilla, Dumm, Dordevic, Homes, Ando, and Basov}}]{AndoRes1}
\bibinfo{author}{\bibfnamefont{Y.~S.} \bibnamefont{Lee}},
  \bibinfo{author}{\bibfnamefont{K.}~\bibnamefont{Segawa}},
  \bibinfo{author}{\bibfnamefont{Z.~Q.} \bibnamefont{Li}},
  \bibinfo{author}{\bibfnamefont{W.~J.} \bibnamefont{Padilla}},
  \bibinfo{author}{\bibfnamefont{M.}~\bibnamefont{Dumm}},
  \bibinfo{author}{\bibfnamefont{S.~V.} \bibnamefont{Dordevic}},
  \bibinfo{author}{\bibfnamefont{C.~C.} \bibnamefont{Homes}},
  \bibinfo{author}{\bibfnamefont{Y.}~\bibnamefont{Ando}}, \bibnamefont{and}
  \bibinfo{author}{\bibfnamefont{D.~N.} \bibnamefont{Basov}},
  \bibinfo{journal}{Phys. Rev. B.} \textbf{\bibinfo{volume}{72}},
  \bibinfo{pages}{054529} (\bibinfo{year}{2005}).

\bibitem[{\citenamefont{Chen et~al.}(1998)\citenamefont{Chen, Kosztin, Jank\'o,
  and Levin}}]{Chen2}
\bibinfo{author}{\bibfnamefont{Q.~J.} \bibnamefont{Chen}},
  \bibinfo{author}{\bibfnamefont{I.}~\bibnamefont{Kosztin}},
  \bibinfo{author}{\bibfnamefont{B.}~\bibnamefont{Jank\'o}}, \bibnamefont{and}
  \bibinfo{author}{\bibfnamefont{K.}~\bibnamefont{Levin}},
  \bibinfo{journal}{Phys. Rev. Lett.} \textbf{\bibinfo{volume}{81}},
  \bibinfo{pages}{4708} (\bibinfo{year}{1998}).

\bibitem[{\citenamefont{Kanigel et~al.}(2006)}]{Kanigelarcs}
\bibinfo{author}{\bibfnamefont{A.}~\bibnamefont{Kanigel}} \bibnamefont{et~al.},
  \bibinfo{journal}{Nature Physics} \textbf{\bibinfo{volume}{2}},
  \bibinfo{pages}{447} (\bibinfo{year}{2006}).

\bibitem[{\citenamefont{Kosztin et~al.}(2000)\citenamefont{Kosztin, Chen, Kao,
  and Levin}}]{Kosztin2}
\bibinfo{author}{\bibfnamefont{I.}~\bibnamefont{Kosztin}},
  \bibinfo{author}{\bibfnamefont{Q.~J.} \bibnamefont{Chen}},
  \bibinfo{author}{\bibfnamefont{Y.-J.} \bibnamefont{Kao}}, \bibnamefont{and}
  \bibinfo{author}{\bibfnamefont{K.}~\bibnamefont{Levin}},
  \bibinfo{journal}{Phys. Rev. B} \textbf{\bibinfo{volume}{61}},
  \bibinfo{pages}{11662} (\bibinfo{year}{2000}).

\bibitem[{\citenamefont{Corson et~al.}(1999)\citenamefont{Corson, Mallozzi,
  Orenstein, Eckstein, and Bozovic}}]{Corson1999}
\bibinfo{author}{\bibfnamefont{J.}~\bibnamefont{Corson}},
  \bibinfo{author}{\bibfnamefont{R.}~\bibnamefont{Mallozzi}},
  \bibinfo{author}{\bibfnamefont{J.}~\bibnamefont{Orenstein}},
  \bibinfo{author}{\bibfnamefont{J.~N.} \bibnamefont{Eckstein}},
  \bibnamefont{and} \bibinfo{author}{\bibfnamefont{I.}~\bibnamefont{Bozovic}},
  \bibinfo{journal}{Nature} \textbf{\bibinfo{volume}{398}},
  \bibinfo{pages}{221} (\bibinfo{year}{1999}).

\end{thebibliography}

\end{document}